\documentclass[3p,twocolumn]{elsarticle}
\usepackage[utf8]{inputenc}
\usepackage{graphicx}
\usepackage{amsmath} 
\usepackage{amssymb}  
\usepackage{amsfonts}

\usepackage{lineno}
\usepackage{xcolor}

\newdefinition{rmk}{Remark}
\journal{Fusion Engineering and Design}

\begin{document}

\begin{frontmatter}
\title{Magnetic control of DTT alternative plasma configurations}
\author[1,2]{E. Acampora\corref{cor1}}
\ead{emilio.acampora@unina.it}

\author[1,3,4]{R. Ambrosino}

\author[5]{A. Castaldo}

\author[1]{R. Iervolino}

\cortext[cor1]{Corresponding author}

\address[1]{{Universit\unexpanded{à} degli Studi di Napoli Federico II, DIETI},
{via Claudio 21},
{80125},
{Napoli},
{Italy}}

\address[2]{{Consorzio RFX, Centro Ricerche Fusione, Universit\unexpanded{à} degli Studi di Padova},
{Corso Stati Uniti, 4},
{35127},
{Padova},
{Italy}}

\address[3]{{DTT S.c. a r.l.},
{Via Enrico Fermi 45},
{I-00044},
{Frascati (RM)},
{Italy}}

\address[4]{{Consorzio CREATE},
{via Claudio 21},
{80125},
{Napoli},
{Italy}}

\address[5]{{ENEA},
{Via Enrico Fermi 45},
{I-00044},
{Frascati (RM)},
{Italy}}

\begin{abstract}

One of the main challenges concerning next generation tokamaks (such as DEMO) will be the development of a heat and power exhaust system able to withstand the large loads expected in the divertor region. A dedicated Divertor Tokamak Test (DTT) facility has been proposed in the EUROfusion Roadmap, with the aim of testing unconventional solutions, such as advanced magnetic configurations and liquid metal divertors. Magnetic control of alternative plasma configurations, such as the \textit{X-Divertor}, will play a key role in the solution of the heat exhaust and yet can be a challenging point, due to increased sensitivity introduced by secondary x-points. 
To overcome the complications introduced by secondary x-points in advanced plasma shapes, magnetic control in DTT is achieved by resolving to the \textit{eXtreme Shape Controller}, in order to control both the plasma shape and the secondary x-point position.
\end{abstract}

\end{frontmatter}

\section{Introduction}\label{section:Introduction}
In 2018, the European Research Roadmap \cite{Eurofusion} set a list of eight missions aimed to tackle the main challenges to the realisation of magnetic confinement fusion. Mission 2 (\textit{Heat-exhaust systems}) calls for an aggressive program aimed to develop alternative solutions for the exhaust of the thermal power of the DEMO Scrape-off layer (SOL), stressing the necessity of a facility dedicated to the study of alternative plasma configurations, exhaust strategies and divertor materials.\\ Following this lead, the DTT (\textit{Divertor Tokamak Test}) project \cite{grbk}, will evaluate the integrability with DEMO of plasma configurations optional to the Single-Null (such as X-Divertor \cite{X-Divertor}, Negative Triangularity \cite{Neg_Triang} and Double-Null \cite{DoubleNull}) and test heat-exhaust strategies such as strike-point sweeping \cite{sweep} and plasma wobbling,  as well as the possibility of a liquid metal divertor.\\
Alternative plasma configurations are advantageous for exhaust control; however, they also introduce complexity both in the design of the tokamak device and in the controllability of the configurations \cite{Reimerdes_2020,MILITELLO2021100908}. 
In this paper, the magnetic control problem of the X-Divertor (XD) configuration for the DTT device is discussed. The XD features a secondary x-point in proximity of the outer vertical target, allowing  to increase flux expansion and connection length and to optimize detachment. However, shape control of XD configurations can be cumbersome, as this secondary x-point also makes the divertor region extremely sensitive.

Control of XD configurations has been already proposed in existing devices like TCV \cite{Reimerdes_TCV_ADC,Degrave}, characterized by a redundant PF coil system, and DIII-D \cite{Kolemen}, with the use of in-vessel divertor coils.
In this paper we propose a generalization of the eXtreme Shape Controller (XSC) \cite{XSC} for the isoflux control of the XD configuration in DTT. The proposed solution, also relevant for DEMO, relies only on the out-vessel coils and could also be generalized to the case of in-vessel divertor coils.\\ Numerical validation has been successfully carried out using the non-linear dynamic simulation code CREATE-NL \cite{CNL}.

 The paper is organized as follows: a general description of the DTT device and of the linearized plasma model is given in \ref{cap1} and \ref{cap21}; in \ref{cap2}, we propose the isoflux surface control solution for the XD configuration; in \ref{cap3}, the XSC is proposed as a strategy of optimal steady state control of the isoflux surface. Finally, closed loop non-liear simulations of plasma behaviour are given in \ref{cap4}, assuming the H-L transition as a possible disturbance.
\section{Poloidal cross-section of DTT} \label{cap1}
\begin{figure}[h]
    \centering
    \includegraphics[width=0.4\textwidth]{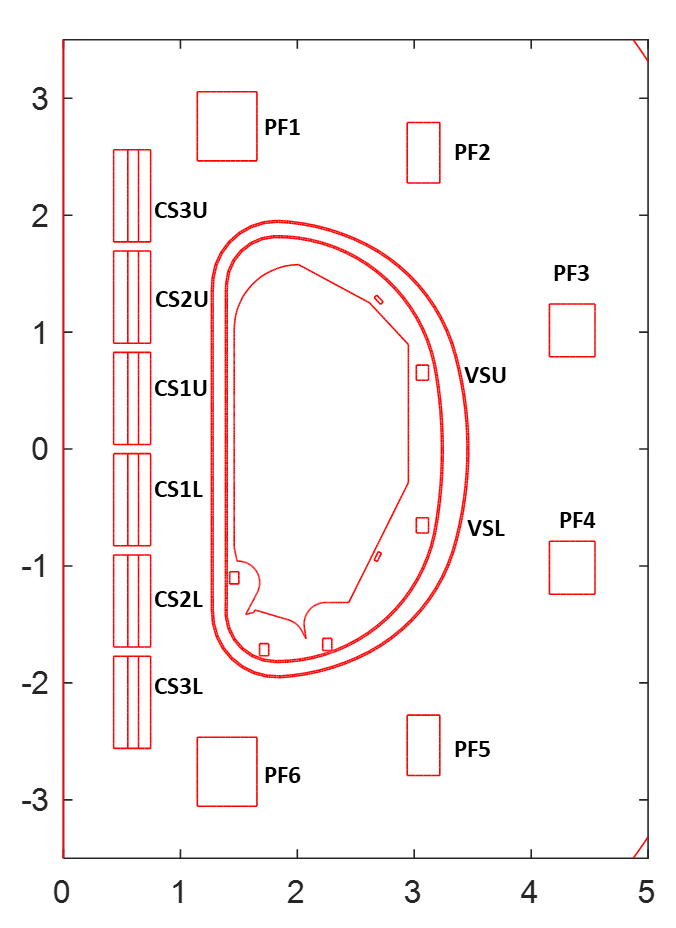}
    \caption{DTT poloidal cross-section.}
    \label{fig:DTTPCS}
\end{figure}

The poloidal cross-section of DTT is reported in Figure \ref{fig:DTTPCS}.
\\
In spite of a major radius of $2.19m$, to maintain a reliable similarity to the challenges of DEMO, DTT is expected to work at a $6T$ toroidal field and a nominal $5.5MA$ plasma current.
A system of six poloidal field coils (PF1-PF6) and six independent Central Solenoid modules (CS3U-CS3L) is responsible for plasma shape and current control. A set of two equatorial in-vessel coils (VSU-VSL) accounts for vertical stabilization (VS) when the coil circuits are connected in antiseries and can also be used for fast radial control when coil circuits are connected in series \cite{AMBROSINO_SOFT} . Two stabilizing plates are foreseen in the upper and lower part of the first wall for passive stability purposes.  Three further in-vessel coils will be used in the divertor region for fine control of the strike points, flux expansion and strike point sweeping \cite{ALBANESE_SOFT},\cite{AMBROSINO2021112330}.
\subsection{DTT linearized plasma model}\label{cap21}
Tokamak devices are usually modeled as an electromagnetic system consisting of the plasma, the passive structures and the active circuits \cite{suite}. The numerical solution to the Grad–Shafranov equation, describing the ideally axisymmetric MHD equilibrium of a plasma in a toroid, is conveyed by the 2D FEM equilibrium code CREATE-NL \cite{CNL}, which outputs a static plasma equilibrium, from which the linearized plasma model can be derived analytically by the CREATE-L code \cite{CL}.\\
The plasma-circuit equation: 
\begin{equation}
   L\dot{I}(t)+RI(t)=u(t)-L_E\dot{w}(t)
   \label{PCE}
\end{equation} where:
\begin{itemize}
    \item $L$ is the mutual inductance matrix among the active coils, the passive structures and the plasma;
    \item $R$ is the resistance matrix;
    \item $L_E$ is the disturbances matrix used to take into account possible
    profile variations;
    \item $I(t)=[I_a^T(t)\;I_p^T(t)\;I_{pl}^T(t)]^T$ is the currents vector, which includes currents on active circuits, eddy currents and plasma current respectively;
    \item $u(t)=[u_{CS/PF}^T(t)\;u_{IC}^T(t)]^T$ is the input vector composed by voltages on CS, PF and in-vessel coils (IC in the following);
    \item $w(t)=[\beta_p(t)\;l_i(t)]^T$ is the disturbances vector, where $\beta_p$ and $l_i$ are measurements of the plasma internal distributions of pressure and current, respectively;
\end{itemize}
leads to the input-state-output form: 
\begin{subequations}\label{plasma_lin_model}
\begin{align} 
    \dot{x}(t)&=Ax(t)+Bu(t)+E\dot{w}(t) \label{plasma_lin_model_A}\\
    y(t)&=Cx(t)+Du(t)+Fw(t) \label{plasma_lin_model_B}
\end{align}
\end{subequations}
 being:
\begin{itemize}
    \item $x(t)=I(t)$;
    \item $A =-L^{-1} R; B = L^{-1}; E =-L^{-1}L_E$;
    \item $y(t)$ the output vector, including measurements such as coil currents, plasma-wall shape descriptors, flux measurements etc.
\end{itemize}

\section{Shape control of the XD configuration}\label{cap2}
The aim of plasma shape control consists in achieving a desired magnetic geometry, in particular for what concerns the Last Closed Flux Surface~(LCFS), in a desired time interval. This is often used to obtain particular magnetic configurations which allow to achieve specific goals, such as improved fusion performances, a better exploitation of the available space or a better distribution of the heat exhaust on dedicated machine structures~\cite{calabro2015east}. 
Shape control of alternative plasma configurations makes this problem even more demanding.
\begin{figure}[h!]
    \centering
    \includegraphics[width=0.35\textwidth]{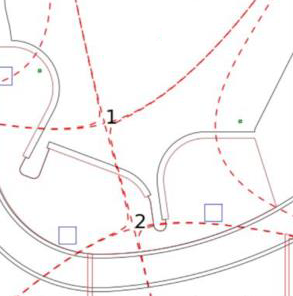}
    \caption{A close view of the DTT X-Divertor concept, showing flux expansion (dashed red) and x-points (numbered 1 and 2) in the divertor region. }
    \label{fig:XD}
\end{figure}
\begin{figure}
    \centering
    \includegraphics[width=0.35\textwidth]{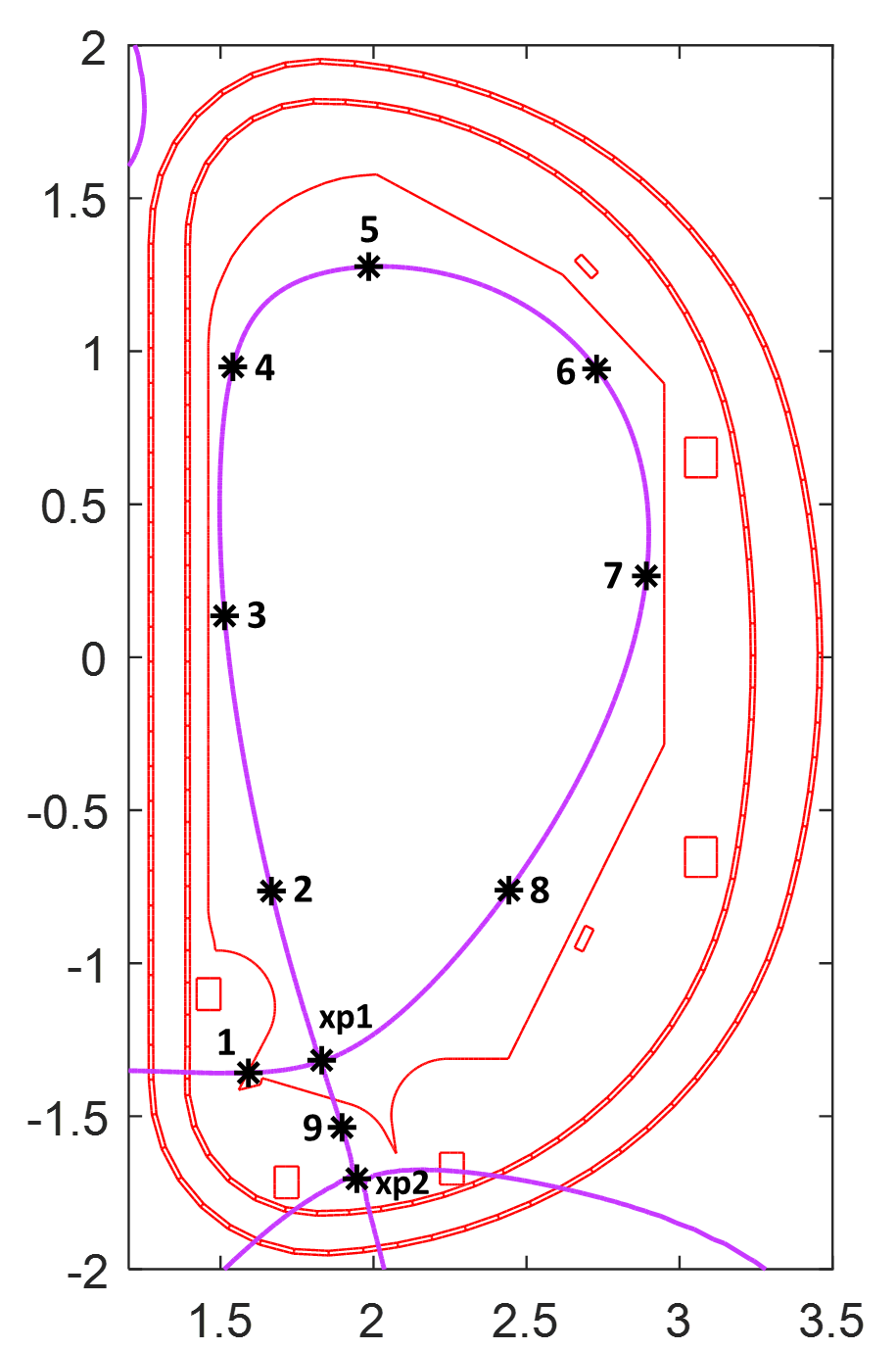}
    \caption{Control points chosen for the isoflux control of the DTT XD configuration.}
    \label{fig:XD2}
\end{figure}\\
In particular, the XD configuration allows to increase the poloidal flux expansion by introducing a secondary x-point behind the divertor target (Figure \ref{fig:XD}).
While the flaring of the flux surfaces allows to increase connection length and optimize detachment, 
the presence of a secondary x-point raises some control problems, such as:
\begin{itemize}
    \item the necessity of additional degrees of freedom for the control of the position and flux of the secondary x-point;
    \item a significant increase of the sensitivity of the plasma shape in the divertor area in case of internal (plasma current profiles) or external (CS/PF current) variations.
\end{itemize} 
The possible approaches for shape control are two: \emph{isoflux} control or \emph{gap} control. The gist of the isoflux approach is to directly control the poloidal flux value at a set of desired boundary locations, forcing a level contour to pass through such points and a chosen x-point (Figure \ref{fig:XD2}). The position of said x-point is also controlled to a target value either directly or controlling the magnetic field components at the null-point location to zero. In gap control, instead, the distance between the LCFS and the first wall along a set of chosen segments is controlled to a desired value. This provides the advantage of a neater physical interpretation of the controlled variables. However, it is possible to rely on gap control only when the reliability of the magnetic field and flux measurements is sufficient to compute the plasma boundary with the required precision. For the case of XD shape control, the isoflux control turns out to be more robust respect to gap control for two main reasons:
\begin{itemize}
    \item the XD flux surfaces in the divertor region are almost flat, making gaps very sensitive to possible variations; 
    \item during transients the shape of the XD risks to move significantly in the divertor region and some of the gaps risk to be not always defined.
\end{itemize} 


 \begin{figure*}[t]
    \centering
    \includegraphics[width=1\textwidth]{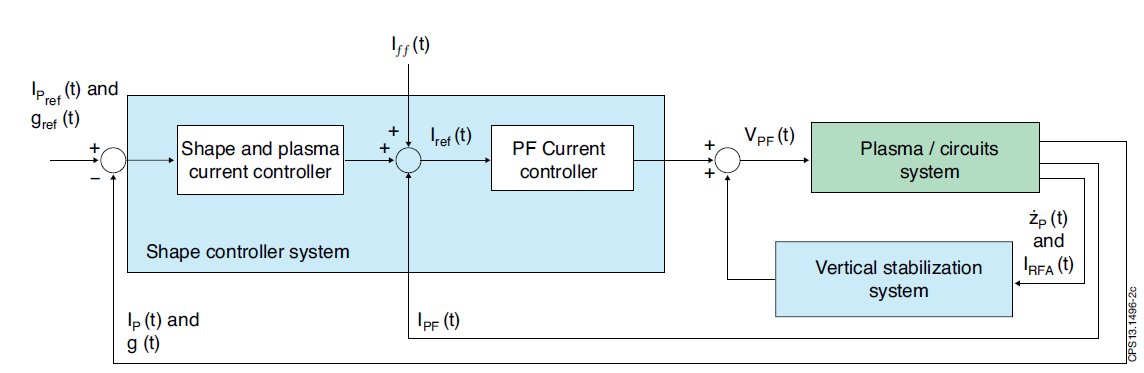}
    \caption{XSC general control scheme}
    \label{fig:scheme}
\end{figure*}

\section{Isoflux eXtreme Shape Control design for the XD}\label{cap3}
\subsection{Controller design}
 The XSC strategy foresees the definition of a \emph{current driven} controller able to generate the current references required by the PF-CS coil system to regulate a given set of geometrical descriptors \cite{XSC}. These references are summed to the feed-forward scenario currents and compared to the current measurements before being fed to a current controller (Figure \ref{fig:scheme}).\\
The controlled shape descriptors $\delta g$ are linked to the PF-CS current variations $\delta I_{PF}$ by: 
\begin{equation}
    \delta g = \Tilde{C}\delta I_{PF}
    \label{pippo}
\end{equation}
where $\Tilde{C}$ consists of the rows and columns of matrix $C$ (Eq.\ref{plasma_lin_model_B}) relative to the shape descriptors and the active coil currents respectively. As $\Tilde{C}$ is generally non-right-invertible, the problem reduces to the identification of the optimal set of currents minimizing the quadratic cost function:
\begin{equation}\label{opt_fun}
    J_1 = \lim_{(t \rightarrow \infty)}(\delta g_{ref}-\delta g)^T(\delta g_{ref}-\delta g).
\end{equation}
In \cite{OptCont} it is shown that the optimal steady state solution to this problem is linked to the inverse of the singular-values matrix of $\Tilde{C}$.
 The XSC has been proved to work well for both plasma-wall distance and isoflux surface shape descriptors \cite{XSC},\cite{adriano}.\\

In the case of the isoflux control of the XD configuration, by indicating with $\bar{sh_i}$ with $i=1,\dots,k$ the $k$ descriptors for the reference shape, $\bar{xp}_1$ and $\bar{xp}_2$ the active and non-active x-points, the following control variables have been chosen:
\begin{equation}
    \delta g = \begin{bmatrix}
    R_{xp_1}\\
    Z_{xp_1}\\
    R_{xp_2}\\
    Z_{xp_2}\\
    \psi_{\bar{sh}_i}-\psi_{\bar{xp}_1}  \\
    \psi_{\bar{xp}_1}-\psi_{\bar{xp}_2}
    \end{bmatrix}
\end{equation}
with $i=1...k$, where:
\begin{itemize}
    \item $R_{xp_1}$ and $Z_{xp_1}$ are the radial and vertical position of the active x-point to be controlled to the reference position
    $R_{\bar{xp}_1}$ and $Z_{\bar{xp}_1}$, respectively.  \item $R_{xp_2}$ and $Z_{xp_2}$ are the radial and vertical position of the non-active x-point to be controlled to the reference position
    $R_{\bar{xp}_2}$ and $Z_{\bar{xp}_2}$, respectively\footnote{Note that, instead of controlling the position of the x-points, it would also be possible to control the radial and vertical poloidal field in the reference x-point positions to zero.}.
    \item $\psi_{\bar{sh}_i}-\psi_{\bar{xp}_1}$ is the difference between the i-flux measurement on the reference shape and the flux in the reference active x-point. The reference value of these control variables is zero.
    \item $\psi_{\bar{xp}_1}-\psi_{\bar{xp}_2}$ is the flux difference between the active and non-active x-points, in the reference position. The reference value of this control variable can be chosen  $>0$ to achieve an XD-minus configuration or $<0$ to achieve an XD-plus configuration. 
\end{itemize}


With this choice:
\begin{equation}
    \Tilde{C}=\begin{bmatrix}
    \tilde{C}_{R_{xp_1}}\\
    \tilde{C}_{Z_{xp_1}}\\
    \tilde{C}_{R_{xp_2}}\\
    \tilde{C}_{Z_{xp_2}}\\
    \tilde{C}_{\psi_{\bar{sh}_i}}-\tilde{C}_{\psi_{\bar{xp}_1}}\\
    \tilde{C}_{\psi_{\bar{xp}_1}}-\tilde{C}_{\psi_{\bar{xp}_2}}
    \end{bmatrix}
\end{equation}
Additional diagonal matrices can be introduced to assign different weights to shape descriptors or to actuators if required:
\begin{equation}
    \Bar{\tilde{C}}=Q\tilde{C}R
\end{equation}
The obtained control action, in the Laplace domain is: 
\begin{equation}
    \delta I_{PF,ref}=PI(s)\Bar{\tilde{C}}^\dagger(\delta G_{ref}(s)-\delta G(s))
\end{equation}
Where $PI(s)$ is a matrix of PI controllers, which can be tuned to adjust the system dynamic response.  
\begin{rmk}
While the XSC is an optimal controller design procedure for the reduction of the tracking errors on the shape descriptors, the value of the weights (and hence of the objective function in \eqref{opt_fun}) and of the PI(s) parameters depends on the specific requirements and constraints.  Their value does not pretend to be optimal but, both in the simulation phase and the experimental activity, they are tuned to improve the control performance avoiding current and voltage saturation of the active coils.
\end{rmk}

\subsection{Estimation of the x-point position}

The implementation of the aforementioned isoflux control strategy requires the real-time evaluation of the flux measurements in the reference control points and the identification of the active and non-active x-points. 
Position and flux measurements in the two x-points can be achieved by recurring to the condition of stationary point for the poloidal flux. A quadratic approximation is assumed for the flux in a given region of the poloidal plane $(r,z)$: 
\begin{equation}
    \psi(r,z)=ar^2+brz+cz^2+dr+ez+f. 
    \label{flux}
\end{equation}
\begin{figure}[h]
    \centering
    \includegraphics[width=0.35\textwidth]{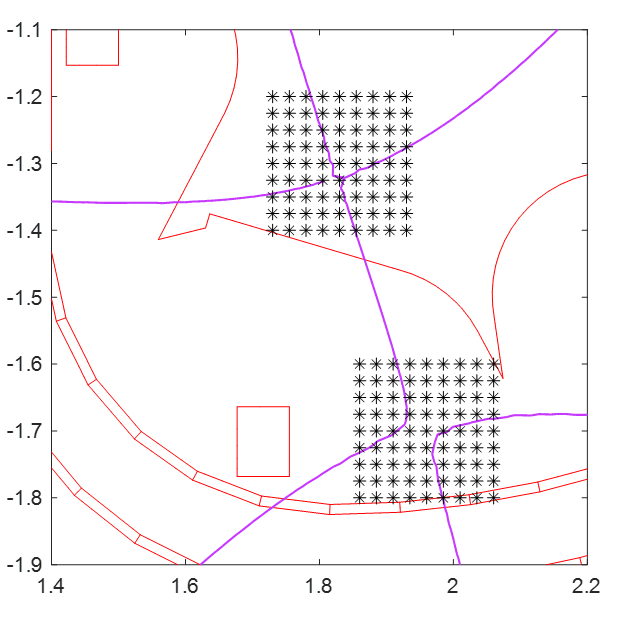}
    \caption{Virtual flux sensor grids chosen for x-point control of the DTT X-Divertor configuration}
    \label{fig:grid}
\end{figure}
If we consider two grids of $n$ virtual flux sensors surrounding the expected x-point positions (Figure \ref{fig:grid}), the relation between flux measurements and sensor positions is known, and coefficients $[a, b, \dots, f]^T$, can be calculated by the Moore-Penrose pseudo-inverse matrix  shown in \cite{suite}. 
The coordinates of the x-points are given by nullifying the gradient of Eq. \ref{flux}:
\begin{equation}
    \begin{bmatrix}
    r_{xp}\\
    z_{xp}\\
    \end{bmatrix}=-
    \begin{bmatrix}
    2a \quad b\\
    b \quad 2c
    \end{bmatrix}^{-1}
    \begin{bmatrix}
    d\\
    e
    \end{bmatrix}
\end{equation}


\section{Closed loop simulations}\label{cap4}
 \begin{figure*}
    \centering
    \includegraphics[width=\textwidth]{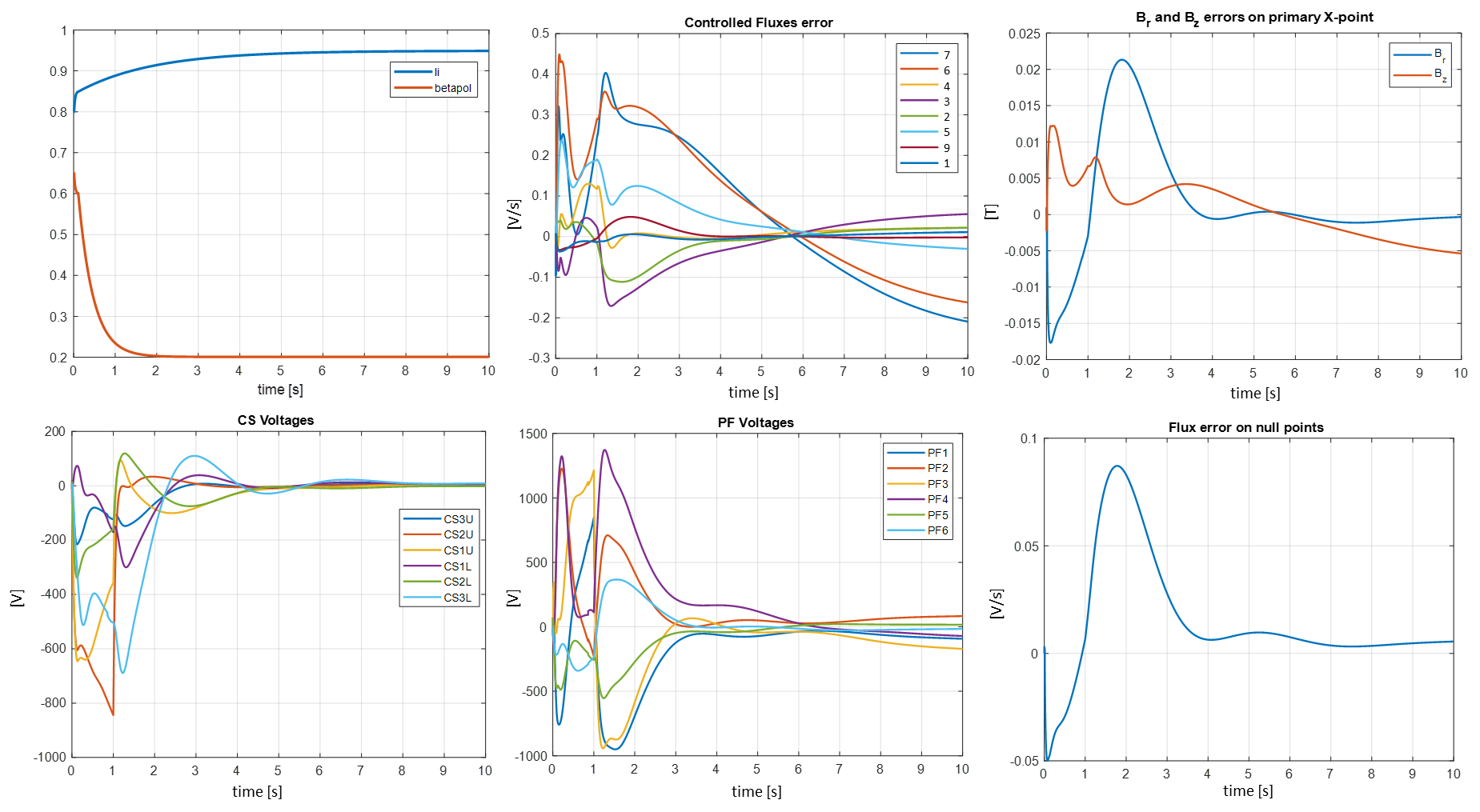}
    \caption{Nonlinear simulation results of DTT isoflux control during an H-L transition.}
    \label{fig:HL}
\end{figure*}
\begin{figure}
    \centering
    \includegraphics[width=0.334\textwidth]{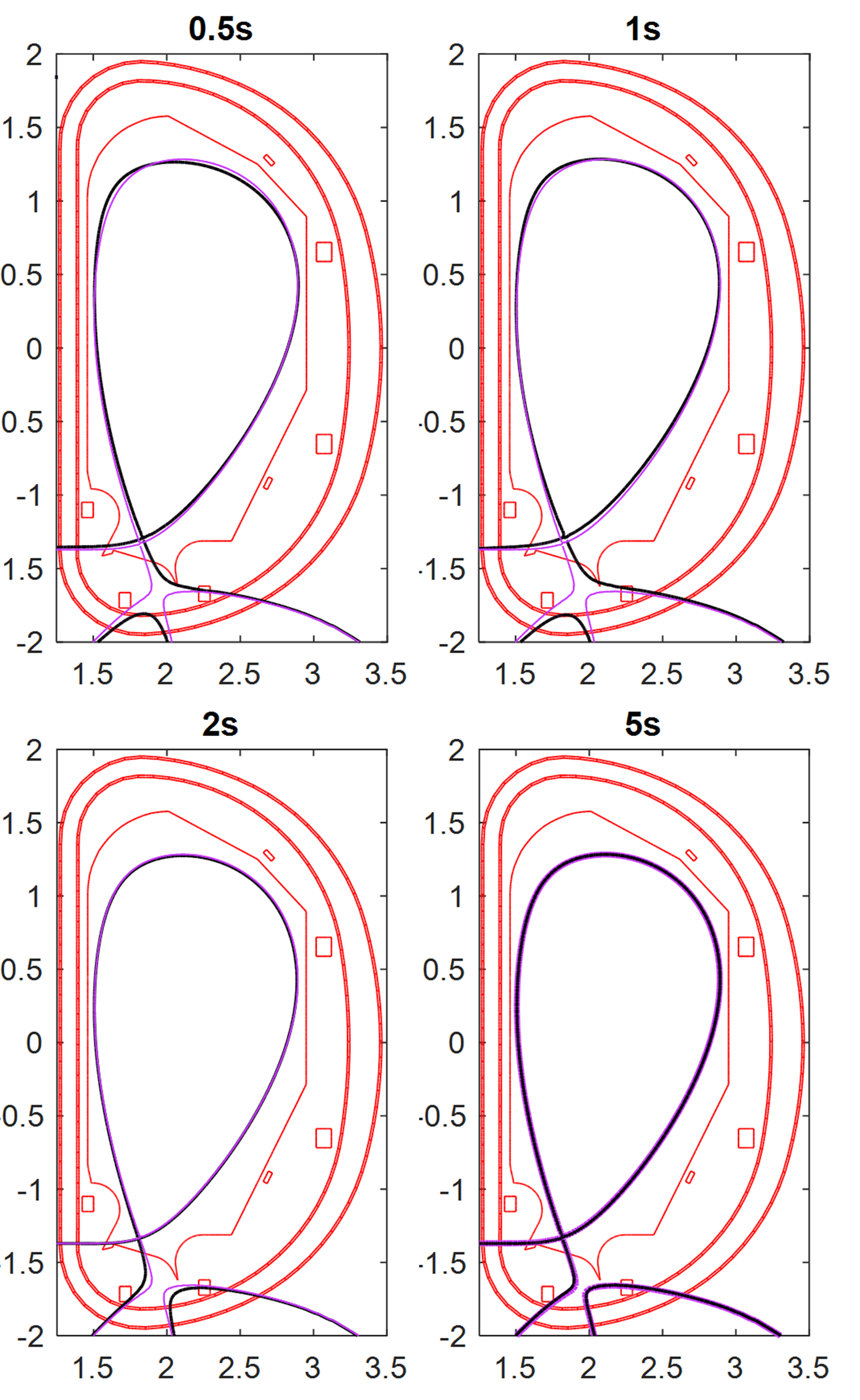}
    \caption{Dynamic evolution of the plasma boundary during the H-L transition. In magenta the target boundary is represented (corresponding also to the initial shape of the plasma before the H-L transition) while the current time boundary is represented in black.}
    \label{fig:sim}
\end{figure}
In this section the simulation results of the closed loop controller are presented.
The controller has been designed using a model based approach; indeed, the MHD equilibrium and the linearized model of the reference DTT XD configuration at flat-top has been produced using the CREATE-L code\cite{CL}. 
The linearized model has been also used for a first test of the control capability and choice of the weights. Finally, the effective validation of the controller has been performed on the dynamic non-linear simulation code CREATE-NL \cite{CNL}.

Concerning the choice of the parameters for the controller, the PI gains are chosen to ensure a time constant of 1s for shape control ($K_p=0.1$, $K_i=4$ on all channels). All fluxes have been weighted of a factor 0.8, while PF5 and PF6 currents have been weighted of a factor 1.25 and 1.67 to avoid the voltage saturation, fixed at $1kV$ for the CS power supplies, $2kV$ for PF1-PF6 and $3kV$ for PF2-PF3-PF4-PF5 \cite{Lampasi_DTT_PS_2022}. All other weights are unitary.\\

In Figures \ref{fig:HL} and \ref{fig:sim} we show the simulation results of the H-L transition, here modeled as a severe exponential $\beta_p$ drop from $0.65$ to $0.1$ in less than $2$s and a rise in $l_i$. The H-L transition represents a very demanding disturbance, causing a fast inner radial movement of the plasma. 
The figures show that the tracking error is kept limited also during the fast transient and the plasma never touches the first wall with the voltage always below the limits. \\
In Figure \ref{fig:sim} it can be noted that, due to the sensitivity of the XD in the divertor region, after 0.5s from the beginning of the H-L transition, the plasma moves from the initial XD-minus in magenta, where the secondary null is included in the low-field side Scrape-Off Layer (SOL), to an XD-plus configuration in black, where the secondary null is contained in the private flux region of the main separatrix. Finally, after 2s, it returns to be an XD-minus, thanks to the closed loop control action.

\section*{Conclusions}\label{section:Conclusions}
An isoflux surface control has been designed for the DTT XD configuration, in the variant of an XSC control. Simulations of the nonlinear plasma model during an H-L transition show that isoflux control can be used to  overcome the controllability complications of alternative magnetic configurations, also during harsh transients.\\
Further developments will imply control validation on the updated DTT asset, which foresees a new divertor with a wide flat dome, new positions of the in-vessel coils and shorter stabilizing plates. 

\bibliography{POD.bib}

\begin{thebibliography}{22}
\expandafter\ifx\csname natexlab\endcsname\relax\def\natexlab#1{#1}\fi
\providecommand{\bibinfo}[2]{#2}
\ifx\xfnm\relax \def\xfnm[#1]{\unskip,\space#1}\fi
\bibitem[{Eur(2018)}]{Eurofusion}
\bibinfo{title}{European research roadmap to the realisation of fusion energy}
  (\bibinfo{year}{2018}).
\bibitem[{Martone et~al.(2019)}]{grbk}
\bibinfo{author}{R.~Martone}, et~al., \bibinfo{title}{DTT Divertor Tokamak Test
  facility Interim Design Report, ENEA, April 2019 ("Green Book")},
  \bibinfo{year}{2019}.
\bibitem[{Kotschenreuther et~al.(2007)}]{X-Divertor}
\bibinfo{author}{M.~Kotschenreuther}, et~al.,
\newblock \bibinfo{title}{On heat loading, novel divertors, and fusion
  reactors},
\newblock \bibinfo{journal}{Physics of Plasmas}  (\bibinfo{year}{2007}).
\bibitem[{Marinoni et~al.(2021)}]{Neg_Triang}
\bibinfo{author}{A.~Marinoni}, et~al.,
\newblock \bibinfo{title}{A brief history of negative triangularity tokamak
  plasmas},
\newblock \bibinfo{journal}{Rev. Mod. Plasma Phys.}  (\bibinfo{year}{2021}).
\bibitem[{Albanese et~al.(2019)}]{DoubleNull}
\bibinfo{author}{R.~Albanese}, et~al.,
\newblock \bibinfo{title}{Electromagnetic analyses of single and double null
  configurations in {DEMO} device},
\newblock \bibinfo{journal}{Fusion Eng. Des.}  (\bibinfo{year}{2019}).
\bibitem[{Ambrosino et~al.(2021)}]{sweep}
\bibinfo{author}{R.~Ambrosino}, et~al.,
\newblock \bibinfo{title}{Sweeping control performance on {DEMO} device},
\newblock \bibinfo{journal}{Fusion Engineering and Design}
  \bibinfo{volume}{171} (\bibinfo{year}{2021}).
\bibitem[{Reimerdes et~al.(2020)}]{Reimerdes_2020}
\bibinfo{author}{H.~Reimerdes}, et~al.,
\newblock \bibinfo{title}{Assessment of alternative divertor configurations as
  an exhaust solution for {DEMO}},
\newblock \bibinfo{journal}{Nuclear Fusion} \bibinfo{volume}{60}
  (\bibinfo{year}{2020}) \bibinfo{pages}{066030}.
\bibitem[{Militello et~al.(2021)}]{MILITELLO2021100908}
\bibinfo{author}{F.~Militello}, et~al.,
\newblock \bibinfo{title}{Preliminary analysis of alternative divertors for
  {DEMO}},
\newblock \bibinfo{journal}{Nuclear Materials and Energy} \bibinfo{volume}{26}
  (\bibinfo{year}{2021}) \bibinfo{pages}{100908}.
\bibitem[{Reimerdes et~al.(2017)}]{Reimerdes_TCV_ADC}
\bibinfo{author}{H.~Reimerdes}, et~al.,
\newblock \bibinfo{title}{{TCV} divertor upgrade for alternative magnetic
  configurations},
\newblock \bibinfo{journal}{Nuclear Materials and Energy} \bibinfo{volume}{12}
  (\bibinfo{year}{2017}).
\bibitem[{Degrave et~al.(2022)}]{Degrave}
\bibinfo{author}{J.~Degrave}, et~al.,
\newblock \bibinfo{title}{Magnetic control of tokamak plasmas through deep
  reinforcement learning},
\newblock \bibinfo{journal}{Nature}  (\bibinfo{year}{2022}).
\bibitem[{Kolemen et~al.(2018)}]{Kolemen}
\bibinfo{author}{E.~Kolemen}, et~al.,
\newblock \bibinfo{title}{Initial development of the {DIII–D} snowflake
  divertor control},
\newblock \bibinfo{journal}{Nucl. Fusion}  (\bibinfo{year}{2018}).
\bibitem[{Albanese et~al.(2005)}]{XSC}
\bibinfo{author}{R.~Albanese}, et~al.,
\newblock \bibinfo{title}{Design, implementation and test of the {XSC} extreme
  shape controller in {JET}},
\newblock \bibinfo{journal}{Fusion Engineering and Design} \bibinfo{volume}{74}
  (\bibinfo{year}{November 2005}) \bibinfo{pages}{627--632}.
\bibitem[{Albanese et~al.(2015)}]{CNL}
\bibinfo{author}{R.~Albanese}, et~al.,
\newblock \bibinfo{title}{{CREATE-NL+}: A robust control-oriented free boundary
  dynamic plasma equilibrium solver},
\newblock \bibinfo{journal}{Fusion Engineering and Design}
  \bibinfo{volume}{96-97} (\bibinfo{year}{2015}) \bibinfo{pages}{664–667}.
\bibitem[{Ambrosino et~al.(2022)}]{AMBROSINO_SOFT}
\bibinfo{author}{R.~Ambrosino}, et~al.,
\newblock \bibinfo{title}{Conceptual design of the {DTT} in-vessel equatorial
  coils},
\newblock \bibinfo{journal}{Fusion Engineering and Design}
  (\bibinfo{year}{2022}).
\bibitem[{Albanese et~al.(2022)}]{ALBANESE_SOFT}
\bibinfo{author}{R.~Albanese}, et~al.,
\newblock \bibinfo{title}{Conceptual design of the {DTT} in-vessel divertor
  coils},
\newblock \bibinfo{journal}{Fusion Engineering and Design}
  (\bibinfo{year}{2022}).
\bibitem[{Ambrosino(2021)}]{AMBROSINO2021112330}
\bibinfo{author}{R.~Ambrosino},
\newblock \bibinfo{title}{{DTT} - divertor tokamak test facility: A testbed for
  {DEMO}},
\newblock \bibinfo{journal}{Fusion Engineering and Design}
  \bibinfo{volume}{167} (\bibinfo{year}{2021}) \bibinfo{pages}{112330}.
\bibitem[{Castaldo et~al.(2018)}]{suite}
\bibinfo{author}{A.~Castaldo}, et~al.,
\newblock \bibinfo{title}{Simulation suite for plasma magnetic control at
  {EAST} tokamak},
\newblock \bibinfo{journal}{Fusion Engineering and Design}
  \bibinfo{volume}{133} (\bibinfo{year}{2018}) \bibinfo{pages}{19--31}.
\bibitem[{Albanese et~al.(1998)}]{CL}
\bibinfo{author}{R.~Albanese}, et~al.,
\newblock \bibinfo{title}{The linearized {CREATE-L} plasma response model for
  the control of current, position and shape in tokamaks},
\newblock \bibinfo{journal}{Nulcear Fusion} \bibinfo{volume}{38}
  (\bibinfo{year}{1998}) \bibinfo{pages}{723}.
\bibitem[{Calabr{\`o} et~al.(2015)}]{calabro2015east}
\bibinfo{author}{G.~Calabr{\`o}}, et~al.,
\newblock \bibinfo{title}{{EAST alternative magnetic configurations: modelling
  and first experiments}},
\newblock \bibinfo{journal}{Nucl. Fus.} \bibinfo{volume}{55}
  (\bibinfo{year}{2015}) \bibinfo{pages}{083005}.
\bibitem[{Ambrosino et~al.(2007)}]{OptCont}
\bibinfo{author}{G.~Ambrosino}, et~al.,
\newblock \bibinfo{title}{Optimal steady-state control for linear
  non-right-invertible systems},
\newblock \bibinfo{journal}{IET Control Theory \& Applications}
  \bibinfo{volume}{1} (\bibinfo{year}{2007}) \bibinfo{pages}{604--610}.
\bibitem[{Mele et~al.(2019)}]{adriano}
\bibinfo{author}{A.~Mele}, et~al.,
\newblock \bibinfo{title}{{MIMO} shape control at the {EAST} tokamak:
  Simulations and experiments},
\newblock \bibinfo{journal}{Fusion Engineering and Design}
  \bibinfo{volume}{146} (\bibinfo{year}{2019}) \bibinfo{pages}{1282–1285}.
\bibitem[{Lampasi et~al.(2022)}]{Lampasi_DTT_PS_2022}
\bibinfo{author}{A.~Lampasi}, et~al.,
\newblock \bibinfo{title}{Power supply systems for the dtt superconducting
  magnets},
\newblock \bibinfo{journal}{IEEE 21st Mediterranean Electrotechnical Conference
  (MELECON)}  (\bibinfo{year}{2022}).

\end{thebibliography}
\bibliographystyle{model1-num-names}
\end{document}